\documentclass{mn2e}
\usepackage{amssymb,graphicx}

\def\etal{{et~al.}}

\def\simlt{\lower.5ex\hbox{$\; \buildrel < \over \sim \;$}}
\def\simgt{\lower.5ex\hbox{$\; \buildrel > \over \sim \;$}}

\title[Have Baryonic Acoustic Oscillations been measured?]
{Have Baryonic Acoustic Oscillations in the galaxy distribution really been measured?}

\author[A. Cabr\'e \& E. Gazta\~naga]{ 
Anna Cabr\'e$^1$, 
Enrique Gazta\~{n}aga$^2$ \\
$^1$Department of Physics and Astronomy, University of Pennsylvania, 209, South 33rd Street, Philadelphia, PA 19104-6396, USA\\
$^2$Instituto de Ciencias del Espacio (IEEC/CSIC), F. de Ciencias, Torre C5- Par-2a, Bellaterra, 08193 Barcelona, Spain.\\
} 
 
\date{Accepted ---. Received ---;in original form ---}

\begin{document} 
 
\maketitle 
 
%\label{firstpage} 

\begin{abstract} 
Recent publications claim that there is no convincing evidence for
measurements of the baryonic acoustic (BAO) feature in galaxy samples
using either monopole or radial information.  Different claims seem
contradictory: data is either not consistent with the BAO model or
data is consistent with both the BAO model and featureless models
without BAO.  We investigate this point with a set of 216
realistic mock galaxy catalogs extracted from MICE7680, one of the largest
volume dark matter simulation run to date, with a volume of 1300 cubical
gigaparsecs.  Our mocks cover similar
volume, densities and bias as the real galaxies and provide 216
realizations of the Lambda or $\omega=-1$ Cold Dark Matter ($\omega$CDM) BAO model.
We find that only $20\%$ of the mocks show a statistically significant
(3 sigma) preference for the true (input) $\omega$CDM BAO model as
compared to a featureless (non-physical) model without BAO. Thus
the volume of current galaxy samples is not yet large
enough to claim that the BAO feature has been detected.
Does this mean that we can not locate the BAO position?  Using
a simple (non optimal) algorithm we show that in 50\% (100\%) of the
mocks we can find the BAO position within 5\% (20\%) of the true
value. These two findings are not in contradiction: the former is
about model selection, the later is about parameter fitting within a
model. We conclude that current monopole and radial BAO
measurements can be used as standard rulers if
we assume $\omega$CDM type of models.
\end{abstract} 
                                   
\begin{keywords} 
galaxies: statistics, cosmology: theory, large-scale structure.
\end{keywords}

\section{Introduction}

%ANNA
Primordial fluctuations generated acoustic waves in the early universe photon-baryon plasma. 
Those waves were frozen at decoupling, $z\sim1100$, then baryon acoustic oscillations (BAO) were imprinted in the 
cosmic microwave background (CMB) 
at the sound horizon scale, as a series of peaks in the power spectrum or a single peak in the 2-point correlation function
(see eg Peebles and Yu, 1970 and Komatsu et al 2010 for the latest
measurements by WMAP).
%EG

BAO can also be seen at the present in matter power spectrum, 
and its position, $r_{BAO}$ can be used as a standard cosmological ruler.
Measurements in the radial (redshift direction), $\Delta z$, can be
used to estimate the Hubble rate as $H(z) =c\Delta z/ r_{BAO}$,
while angular measurements, $\Delta \theta$, can be used
to estimate the angular diameter distance: $D_A(z) =r_{BAO}/ \Delta \theta$.
%EG
Baryon acoustic oscillations in the galaxy correlations of
the Sloan Digital Sky  Survey (SDSS) luminous red galaxy (LRG) 
sample have been used to constrain cosmological parameters 
(eg Eisenstein et al 2005, Hutsi et
al 2006, Sanchez et al 2009,  Percival et al 2010, 
Reid et al 2010, Kazin et al 2010a and references
therein). Different studies use different ways to extract the BAO signal
and quantify the significance of the measurements (see Sanchez et al
2008). 
For example,  Eisenstein et al 2005 and Sanchez et al 2009 used the
full shape of the  2-point correlation to $\omega$CDM class of models 
and found constraints to
the combination distance  $D_v(z)=(D_A^2/H)^{1/3}$
to the galaxy sample mean redshift based on a global $\chi^2$
fitting, while Percival et al 2010 used a fit to the oscillatory components
in the power spectrum to find constraints on $D_v(z)$.

These previous analysis used the monopole component of the correlation
function, where all pairs are averaged with independence of their
orientation. Okumura et al 2008 did a separate analysis of pairs 
as a function of orientation but avoiding the radial direction.
Gaztanaga, Cabr\'e and Hui (2009, GCH hereafter) presented 
constraints to $H(z)$ based
on  the radial correlation, which uses only those pairs aligned with the
redshift direction. 
%ANNA
This reduces the number of observational data but boost the contrast on 
the BAO peak because of redshift space distortions.
At intermediate scales, lower than BAO, the correlation function
becomes negative  in the line-of-sight direction, creating a better 
contrast in the BAO position, easier to detect 
than in real space.  Also, non-linearities, magnification and bias can boost the 
peak (see GCH and Tian et al 2010 for further details).

GCH presented two ways to analyze the BAO data: the
peak and the shape method. In the peak method they find the location
of the peak and use it as standard ruler to measure $H(z)$. In the shape method they use
a $\chi^2$ fit to the full shape of the correlation and find the best shift in the distance
$H(z)/H_0$. The shape method was also  used to test if the data was
compatible with the shape of the correlation expected in $\omega$CDM. They compare
different classes of models: the standard BAO $\omega$CDM model,
 a similar class of models without BAO (so called no-wiggle 
model in Eisenstein \&
Hu 1998) and a model with zero correlation $\xi=0$. The no-BAO
model has $\Delta \chi^2=10$ with respect to the best fitting $\omega$CDM model
while a model with $\xi=0$ has  $\Delta \chi^2=4$.
Kazin et al (2010b)  did an independent analysis of the SDSS catalog 
and found similar results for the correlation measurements and errors.
In their interpretation they did not explore the parameter space
of $\omega$CDM but conclude that there is no convincing evidence 
for radial BAO
because the $\xi=0$ model fit the data better than $\omega$CDM.
They argue that there are no parameters in the $\xi=0$ model
while for $\omega$CDM several parameters where fitted in GCH.
After including the penalty for adding parameters, 
they find that $\omega$CDM is not significantly better than  $\xi=0$. 

But a similar  argument could be extended to the BAO monopole
measurements. 
For example, if one fits a constant correlation to
the LRG correlation function in Fig.17 of Sanchez etal (2009) to scales
larger than 70 Mpc/h one finds that this model can not be
distinguished from a $\omega$CDM model with free parameters. The original
Eisenstein etal (2005) results can also be well fitted with a
power-law  model$^1$.
Does this mean that the BAO feature has not been
detected at all?
These are important points to clarify as it is common
practice to include BAO measurements when fitting cosmological models
to provide evidence for dark energy models (eg 
Sanchez et al 2009; Komatsu et al 2010; Kazin et al 2010a;
Gaztanaga, Miquel \& Sanchez 2009).

Other recent studies seem to reach a similar conclusion, that the BAO feature 
has not been detected, but using an argument that seems to
go in the opposite direction. Rather than finding that data is too noisy
and compatible with featureless  models, they find that the data
is not consistent with $\omega$CDM 
%due to an excess power at large scales,  above 140Mpc/h
(eg see Labini et al 2009, Labatie et al 2010). 
Also see Martinez et al 2009 for a study of peak detection using DR7 monopole.
We will investigate this here to find, as in previous analysis (eg
GCH, Sanchez et al 2009, Kazin et al 2010a) that data is in good agreement with 
$\omega$CDM although we should stress that this statement will depend
on the specific test we use.

 We will argue that there are two separate
questions mixed up in the above line of argumentation: 
model selection and parameter fitting. We will find that 
while current data can not be used to select $\omega$CDM,
one can still constrain the parameters of $\omega$CDM if this model
is assumed. To show this, we will set out to address two main questions:
1) can we use current BAO data to favor $\omega$CDM? In other words: 
is the volume of current data large enough to pass a null detection
test to choose $\omega$CDM over some other model?
2) can we  constrain the parameters of the $\omega$CDM model,
and in particular the BAO position with current data?

We will investigate these points  with a
set of 216 mock galaxy catalogs extracted from MICE7680
(see Fosalba et al 2008, Crocce et al 2009),
one of the largest volume dark matter simulation run to date.
The mocks are made to match the SDSS LRG DR6 sample and should
therefore provide a good representation of biased $\omega$CDM realizations.
We will use these mocks to explore the peak and
the shape method applied to the monopole. We use the monopole here
(rather than radial BAO) for several reasons: shape measurements have
larger  signal-to-noise, theoretical modeling of monopole is better
understood (see GCH) and the monopole BAO has been more widely used to
test cosmological models.
Rather than comparing the $\omega$CDM with some add-hoc correlation
(power-law, constant or some combination)
we choose to focus on comparing BAO and no-BAO models.
This has the advantage of being a well defined procedure 
(quite standard in the literature) where we
have the same number of parameters in each case, which simplifies
the interpretation of the statistical significance when comparing two
different models with different number of parameters (eg see Liddle 2009).

Throughout we assume a standard cosmological model, with 
$\Omega_{\rm M}=0.25$,$\Omega_{\Lambda}=0.75$, $\Omega_{\rm b}=0.044$,
$n_s=0.95$, $\sigma_8=0.8$ and  
$h\equiv H_{0}/(100\,{\rm km\,s}^{-1}{\rm Mpc}^{-1})=0.7$.

\section{BAO in galaxy mocks}

\begin{figure}
\includegraphics[width=80mm]{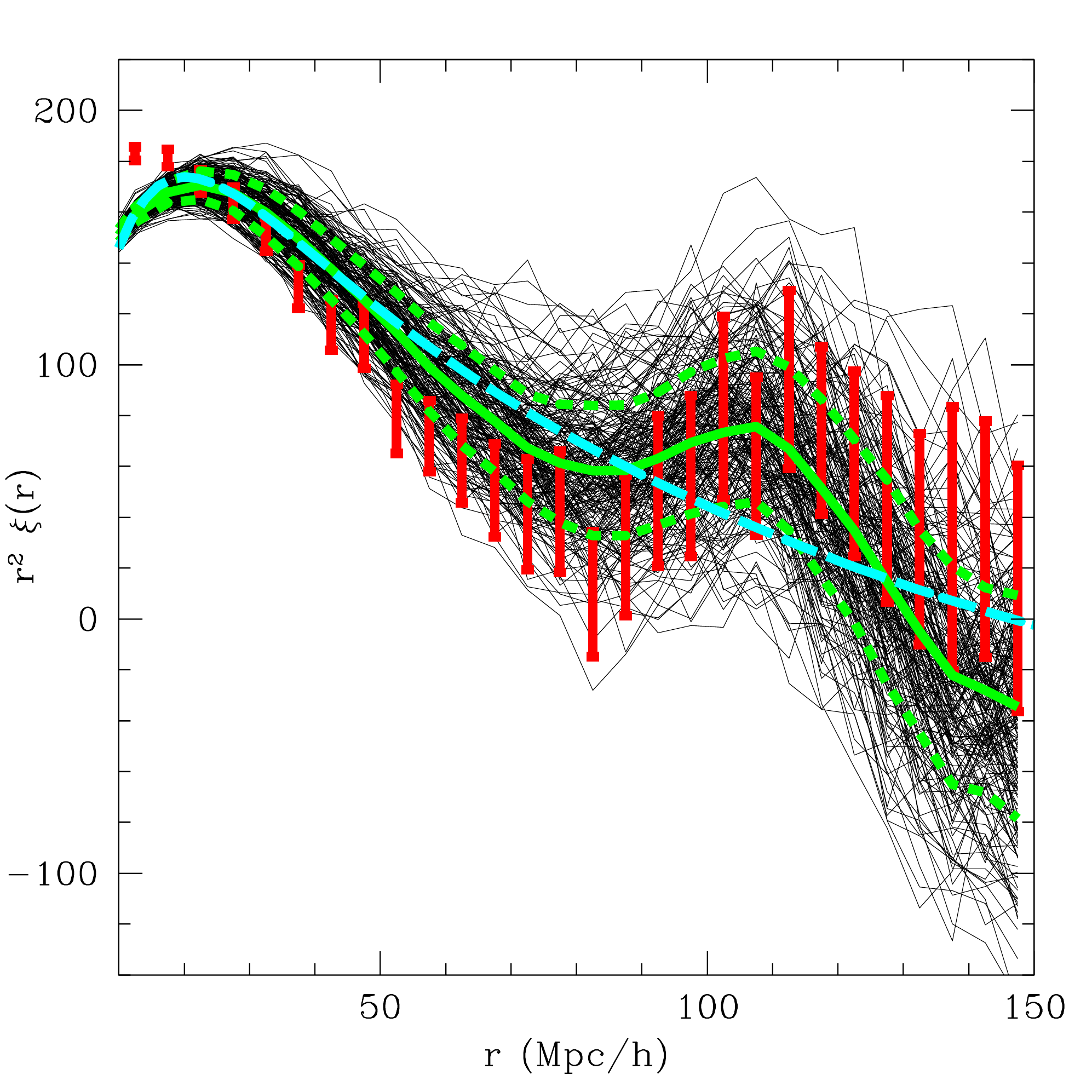}
%\plotone{figs/x2models.pdf}
\caption
{Thin (black) lines show the correlation function $\xi(r)$ (scaled by $r^2$)
 in each of our 216 mocks.
Solid (green) line shows the BAO model (the mean of the mocks).
Short dashed (green) lines encompass 1-sigma errorbars from the mean 
(these are mock to mock errors, 
the error of the mean would be $1/\sqrt{216}$ better).
Long dashed (blue) line shows the no-BAO model.
The (red) errorbars correspond to the real LRG data shifted
as shown in Eq.\ref{eq:scale}.}
\label{fig:correlation}
\end{figure}

Appendix A in Cabr\'e \& Gaztanaga 2009 (CG09 from now on)
describes how our mocks were
built and also how the correlation function is estimated.$^1$
We include both bias and redshift distortions in the mocks.
We will focus here on the monopole correlation for halo z=0 mocks with
a bias $b\simeq 2$, similar to LRG galaxies.
The correlation function for our 216 mocks, its mean and errors
are displayed  in Fig.\ref{fig:correlation}.
These mocks are realistic as
they cover similar volume, densities and bias as the real LRG
galaxies, but they have some limitations.
In general, one needs first to explore the parameters in $\omega$CDM (and bias model) to get
a good match to data. Our simulations have $\beta\equiv
f(\Omega_m)/b \simeq 0.25$ and $z=0$, which are different from the values
in real data $\beta \simeq 0.34 \pm 0.03$ and $z=0.35$ 
(the difference in $\beta$ comes from the difference in
redshift, as bias is similar, see CG09). Depending on the test used,
this could result in a poor fit of models to data.
Despite these limitations, we will find
below a good fit of data to the mocks when we allow the amplitude to
vary in the fit. This  indicates that our mocks
provide a good representation of the data, given the errors,
at least for the questions we want to address here.

In our analysis we will pretend that each mock is a realization
of the real LRG data. 
Our mocks are close enough to the real data
to provide a realistic representation of how much variation there is
from one realization of real data to the other. Indeed the jack-knife
(JK) errors (and covariance matrix)  in the real data 
are similar to the JK errors in our mocks
and to the ensemble variation from mock to mock. This 
was shown in CG09 and can also be
seen in Fig.\ref{fig:correlation} where we compare the ensemble
variation in mocks (short dashed lines)
to the JK errors in  the DR6 SDSS LRG measurements from CG09 (note
that we show DR6 to be consistent with the mocks, but similar
results are found for DR7, see GCH).
To compare to simulations we have scaled the LRG data as:
\begin{equation}
\xi(r) \rightarrow A \left[\xi(r) +K \right]
\label{eq:scale}
\end{equation}
with $A=1.2$ and $K=-0.005$. 
%ANNA
The value of $A$ accounts for the differences 
between the simulation and LRG data in $\beta$, growth and bias.
The value of $K$ represents a possible, but quite
minor ($0.25\%$),  error (contamination or sampling fluctuation) in the overall mean 
density of the sample. This has  little impact in the fit of models
to data (covariance allows for a constant shift in the data)
but improves the visual comparison in the figure
(see Fig.17 in Sanchez et al 2009). 
As indicated by Fig.\ref{fig:correlation}  the mocks represent quite well 
the variation seen in the observational data. 
%CG09 used 3 different sets of 216  mocks each to
%encompass the actual values of $\beta $ and $z$ in the real data.
%We find very similar results when we use these other mocks in the
%analysis below and are therefore confident that the results presented
%here are quite generic.

\subsection{The shape method: null test}

We  use two models to fit the correlation $\xi(r)$:
1) {\it the BAO model}: it uses the mean of all the mocks in order to have a
perfect  BAO model (with bias, redshift space and non-linearities effects included).
2) {\it the no-BAO model}: a non-physical model that imitates well the broad
band correlation but does not include a BAO peak. 
We use the no-wiggle power spectrum of Eisenstein\& Hu (2001)
with same $\omega$CDM parameters as the simulation. 
%Note that the later model
%is in real space and in linear theory (both these effects are small,
%given the errors and the fact that we will fit the amplitude).
Fig.\ref{fig:correlation} compares the BAO  
(solid line) with the no-BAO model (long-dashed  line).
Our null test is: does the data prefer
the BAO to the no-BAO model at 3-sigma confidence level (CL)?

To simplify the analysis and interpretation,
the only free parameter that we fit is the global amplitude $A$ 
of the correlation, which includes a possible bias (as we are using halos) and 
a constant redshift distortion boost (Kaiser 1987).
We use the correlation function $\xi_i(r_j)$ measured in the $i$-th
 mock at separation $r_j$ to perform a $\chi^2$ fit and find the best fit amplitude 
$A_i$ for either BAO or no-BAO models (which are labeled generically
as $\xi_m$):
\begin{equation}
\chi^2_i = \sum_{jk} \left[\xi_i(r_j)-A_i \xi_{m}(r_j)\right]
C_{jk}^{-1} 
\left[ \xi_i(r_k)-A_i\xi_{m}(r_k) \right]
\end{equation}
The indexes $j$ and $k$ run over the $N_b=20$  bin separations, ie
$\nu=19$ degrees of freedom.
Bins are linearly spaced with $\Delta r=5$ Mpc/h between 30 and 130
Mpc/h (we find similar results in the range 20-150 Mpc/h).
The covariance matrix $C_{jk}$ is estimated from the mocks:
\begin{equation}
C_{jk}= {1\over{215}} \sum_{i=1}^{216}  \left[\xi_i(r_j)- \bar{\xi}(r_j)\right]
\left[\xi_i(r_k)- \bar{\xi}(r_k)\right]
\end{equation}
where $\bar{\xi}(r_j) \equiv {1\over{216}} \sum_i \xi_i(r_j)$ is the mean value in bin $j$.

\begin{figure}
\includegraphics[width=60mm]{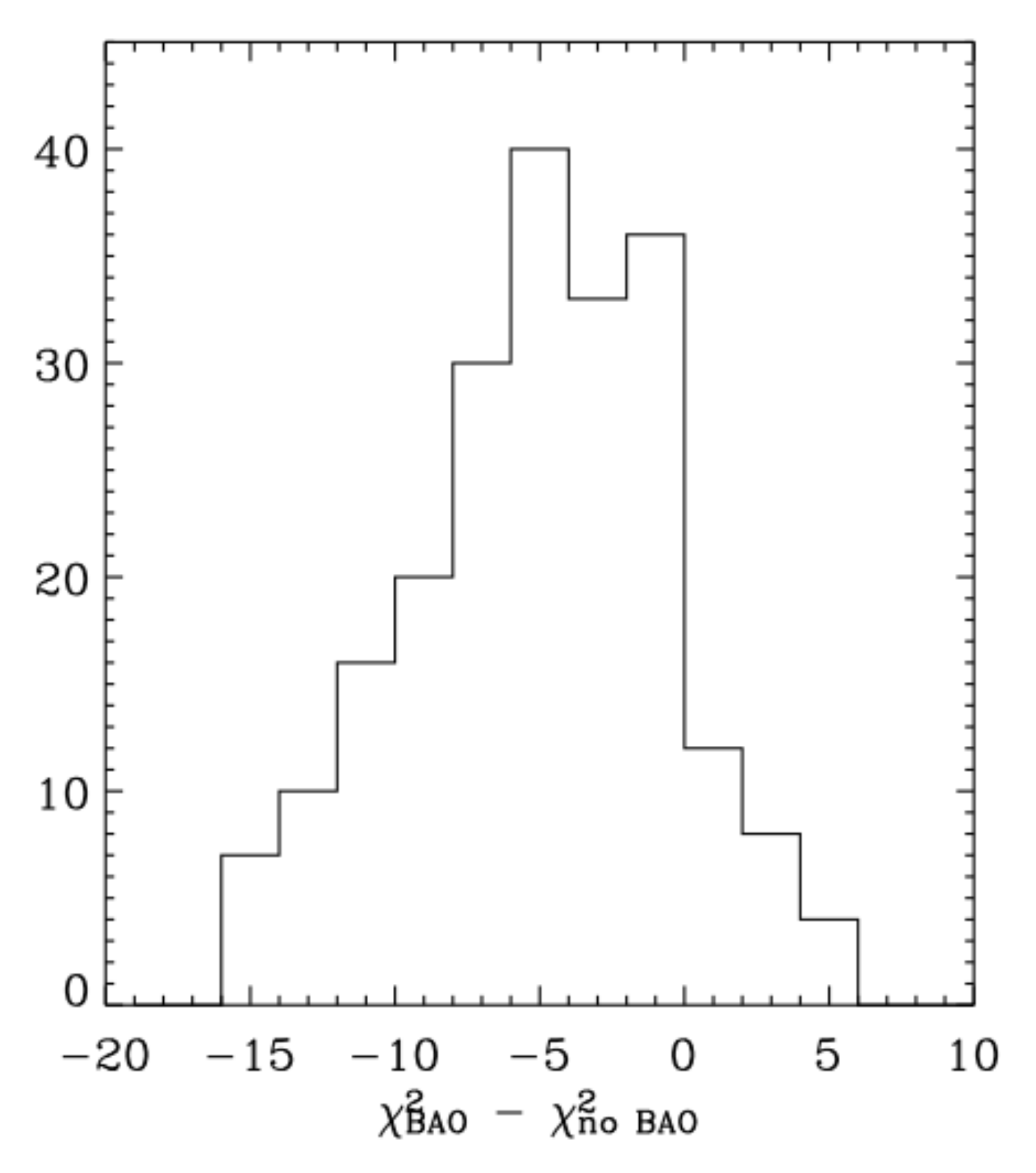}
%\plotone{figs/chi2.pdf}
%\plotone{figs/dchi2.pdf}
\caption
{ Histograms showing the distribution of differences in $\chi^2$ values
for  different LRG mocks.
The correlation function in each mock is fitted with both the standard $\omega$CDM
BAO  correlation and with the
no-BAO class of models. The difference between the two $\chi^2$ values
in each mock is accumulated in the histogram. There are $N_b=20$ 
data bins in each fit, but only one parameter is fitted (the overall amplitude).
The figure shows that only $20\%$ of the cases show
a significant preference at $3\sigma$ (ie $\Delta\chi^2<-9$)
for the BAO model over the no-BAO model. The mean for mocks is $\Delta\chi^2=-5$.
A fit to the real LRG data gives also $\Delta \chi^2=-5$, close
to the maximum.}
\label{fig:dchi2}
\end{figure}

The resulting distribution of values of $\chi^2_i$ for the BAO model
peaks around $\chi^2_i \simeq \nu = 19$ and is quite broad ($\Delta\chi^2
\simeq \sqrt{2\nu} \simeq 6$, as expected).  
The no-BAO model peaks at larger values ($\chi^2_i
\simeq 24$) and is  slightly broader ($\Delta\chi^2 \simeq 7.7$). 
The real LRG data produces
$\chi^2=20$ for the BAO model and $\chi^2=25$ for the no-BAO model,
well within the values found for most of the mocks.  Thus, given the
large errorbars, the real data seems to match quite well our mocks,
despite the differences in the modeled values of $\beta$, bias and $z$
mentioned above. 

In Fig.\ref{fig:dchi2} we plot the histogram of the differences
between the $\chi^2_i$ values in the fits to the BAO and
no-BAO models for each mock. Negative values mean that the mock prefers
the BAO model over no-BAO model.
A difference  at $3\sigma$  CL
between both models, ie $\Delta\chi^2 < -9$, 
only happens in the 20\% of cases (up to 30\% when we
explore other range of scales). This means than in $80\%$ of
the cases one does not expect to be able to distinguish between the
two models (at more than $3\sigma$ CL). This result is not surprising.
The mean difference in $\chi^2$ between the BAO and no-BAO model is only
$\Delta\chi^2 \simeq -5$, which in comparable to the width of the
$\chi^2$ distribution with 19 degrees of freedom.
In other words, current
errors  are still  too large  to claim a BAO detection.
%Does this mean that the BAO is not present in the mocks or data?

\subsection{The peak method: BAO position}

In the peak method, we assume that we live in a $\omega$CDM universe and
try to locate the BAO position. To keep things simple,
here we locate the position of the peak by searching the maximum in 
the correlation function in the BAO scale, between 80-135Mpc/h
(results are similar when we move around 70-150Mpc/h). 
The BAO feature is modified by the presence of the 
broad band (CDM) correlation function, which  can be modeled approximately by a power
law.  We fit a power law to each correlation function at small scales (10 - 70Mpc/h)
and subtract the correlation function from the best power law before
locating the peak. 
GCH use a very similar peak method but
do not need to subtract the power-law because the correlation is
quite flat (and close to zero) in the radial direction.
Sanchez et al (2010) use a similar but more elaborated version, where
they fit simultaneously a power-law,  a constant shift and a gaussian
(BAO) peak. This would provide more accurate errors for the peak.
%Our goal here is not so much to find the best way to locate the BAO peak but
%rather to just show that the peak is located in a significant way.

\begin{figure}
\includegraphics[width=60mm]{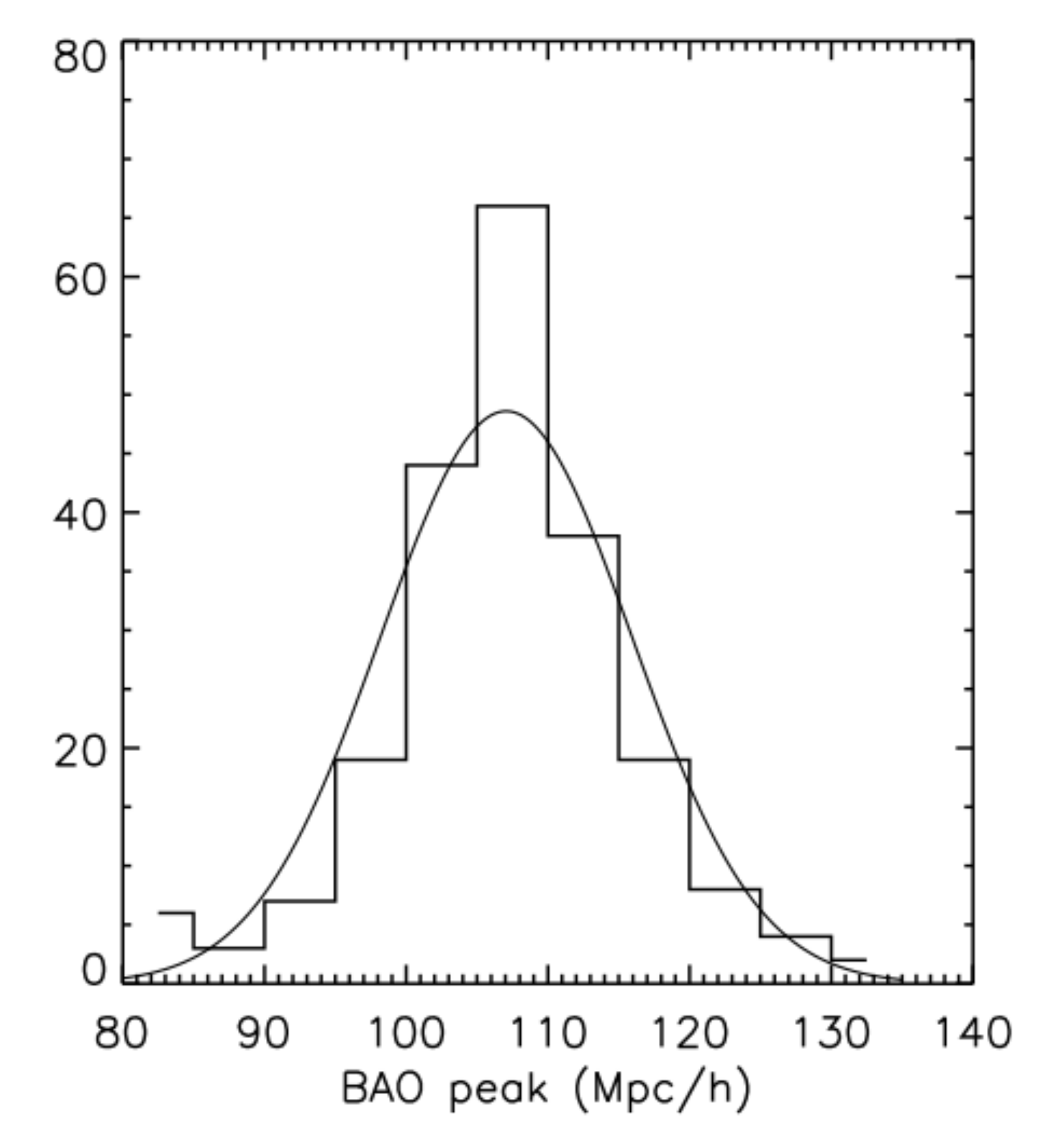}
%\plotone{figs/hbao.pdf}
\caption
{Histogram showing the distribution of peak BAO position measured in LRG mocks.
This distribution has $r_{BAO}= 107.2 \pm 8.8$ Mpc/h as compared to
$r_{BAO}=107.5$ Mpc/h in the mean model. The distribution
is quite gaussian, as shown by the line crossing the
histograms. The same measurement in real LRG DR6
data yields $r_{BAO}=112$ Mpc/h.}
\label{fig:hbao}
\end{figure}

Fig. \ref{fig:hbao} shows the distribution of recovered
BAO positions for individual mocks. This distribution
is well approximated by a Gaussian (also shown in the figure).
%ANNA
The mean BAO position is 107.2Mpc/h with a dispersion of
8.8Mpc/h,  compared to mocks mean value position at 107.5Mpc/h (note that
the resolution in the position of the peak is of 5Mpc/h).
The position of the peak can differ slightly (less than 2\%) from the sound horizon scale at 
decoupling (see S\'anchez, Baugh and Angulo 2008 and S\'anchez et al 2010) depending
on the cosmology, non-linearities, and other effects. 
For WMAP parameters, very similar to MICE simulation, the sound
horizon scale is at 107.3Mpc/h. 

We find similar results when we use a fixed power-law for all mocks,
the one fitted for the mean correlation function,
or when using the best power-law for each mock,
as one would do in real data. 
When we apply the same method to the
real DR6 data we find  $r_{BAO} =112Mpc/h$,  well within the bulk
of our mocks. 

We have also tried the method to locate the BAO position
propposed by Kazin et al (2010a). The mocks
(or data) are fitted using a $\chi^2$ likelihood (including covariance)
to a BAO model which consists in the mean of the mocks, $\xi_m(r)$, shifted 
by two free parameters: the amplitude $A$ and a scale shift $\alpha$, ie 
$A\xi_m(\alpha r)$. We find a very similar histogram to that in Fig.\ref{fig:hbao}
but with smaller errorbar: $6\%$ instead of $8\%$.
Kazin et al (2010a) further reduced this error to $3-4\%$  by using
only the mocks which have a clear BAO as in the DR6 data.
%EG
We will obviusly get smaller errors by removing such outliers in our mocks
but this later step involves more assumptions than just the existance
of a peak. It not only assumes that we  live in  $\omega$CDM, but
selects in a subjective way (a posteriori) within a subset of realizations. 
%For the DR6 data we find that this second method to locate the peak yields 
%$r_{BAO} =101Mpc/h$ for 80-140 Mpc/h when using the mean of all the
%mocks.  But this result is very
%EG
%ANNA
%sensitive to the range of scales and small changes in the data. 
%We think that this
%could partially reflect noise and may be also the fact that our mocks are not realistic
%in detail (ie different values of $\beta$ and of mean redshift), so 
%a fixed shape based on these mocks is not working properly with SDSS, 
%which might not have exactly the same parameters than simulations.
Also note that in this method we are using a priori knowledge of the shape of the
input model to locate the peak. In the peak method, used in Fig.3, we do not
need to make such assumption and so we think this makes a stronger
case for the point we want to demostrate, even when the error is larger.

It is more robust and self-consistent to locate the BAO and error 
using the full shape of $\xi(r)$ and a larger family of
cosmological models, eg as shown in Sanchez et al (2009), 
% ANNA
avoiding any dependence on a particular cosmology in the algorithm to locate the peak.
The point demonstrated here is that the BAO
position is imprinted in the mocks despite the
fact that they  do not pass a null
detection test. 
%ANNA
A comparison between methods is left for future analysis.

\section{Conclusion}
\label{sec:conclusion}

The first question we set out to address was if the volume of current 
BAO data is large enough to pass a null detection test for $\omega$CDM.
The answer to this question seems negative.
We have shown in Fig.2 that the distribution of
$\chi^2$ differences is quite broad and one could find mocks for which the null
test is passed or failed. In fact $80\%$ of the mocks have
$\Delta\chi^2>-9$, which indicates
no statistically significant (at 3-sigma CL) preference for the true BAO
input model as compared to the featureless no-BAO family.
 Current SDSS (DR6-DR7) data seems to lie close to the peak of this distribution,
$\Delta \chi^2 \simeq -5$, but according to Fig.2
this does not provide convincing evidence for the BAO model. 
As expected, the DR3 results in Eisenstein et al 2005 (about
half of the DR6 volume) is even less significant: $\Delta \chi^2 \simeq
-1.1$. When we compare the BAO model to a power-law fit (with 2 parameters) we find
$\chi^2_{BAO}-\chi^2_{power-law}  \simeq 0.4$.$^1$ 

Our mocks have slightly different values of $\beta$ and $z$ than the
DR6 data (see Fig.1) and we wonder if this could affect the above conclusion.
The important point to notice is that the BAO and no-BAO model also
have a similar difference of $\Delta\chi^2$ (ie $\simeq -5$) 
when we compare to the DR6 data
(using the same bins and covariance as in the mocks). 
Models with other cosmological parameters 
within the uncertainties of $\omega$CDM also produce similar
$\Delta\chi^2$.  But a value of $\Delta \chi^2 \simeq -5$ is  
comparable to the width of a $\chi^2$ distribution with 19 degrees 
of freedom (which is $\Delta\chi^2 \simeq 6$). 
This is why the result is not significant.
Our conclusion is quite robust and mostly relays in the size of the
errors  and covariance between
bins. The covariance estimate is consistent in the data 
(ie from Jack-knife subsamples), in our mocks
and in other mocks produced by several groups (eg Eisenstein et al 2005,
CG2009, Kazin et al 2010a). We would need  $\sqrt{1.8}$  times smaller errors, 
ie 1.8 times more data or some optimal weighting
(Hamaus et al. 2010, Cai etal 2010),
than DR6 (so that $\Delta\chi^2$ increases from 5 to 9)
to be able to claim a 3 sigma BAO detection in the monopole. 

%ANNA
%For the radial BAO analysis, GCH find  $\Delta\chi^2 \simeq -10$ in DR6
%for 20 degrees of freedom (5Mpc/h radial bins within 40-140 Mpc/h)
%when comparing BAO and no-BAO models. This seems more
%significant than the monopole, probably because the radial BAO
%peak is boosted by redshift space distortions (and maybe magnification bias).
%A model with $\xi=0$ has only  $\Delta \chi^2=-4$, which has been
%used to question this detection (Kazin et al 2010b).

For the radial BAO analysis, GCH reach similar conclusions. 
They find that the difference  $\Delta\chi^2 \simeq -10$ in DR6                     
for 20 degrees of freedom (5Mpc/h radial bins within 40-140 Mpc/h)
when comparing BAO (plus magnification) and no-BAO models
(without magnification the difference is $\Delta\chi^2 \simeq -6$).
%EG
This seems more significant
than the monopole, probably because the radial BAO                                                                        
peak is boosted by redshift space distortions and magnification.

Does this mean that the BAO position can't be measured?
If we assume the $\omega$CDM model,
we can locate the BAO position to better than $8\%$ of the true
value,  as illustrated in Fig.3. 
% ANNA
We show that in 50\% (100\%) of the mocks we can find the BAO position
within 5\% (20\%) of the true value.
This error is an upper bound as we
have not tried to optimize the method to locate the peak.
We have compared the mocks with the real data and
found no evidence for deviations away from the $\omega$CDM.
None of the 216 $\omega$CDM  realizations is identical
to the measurements (or in fact to each other), 
but observations produce values that lie well within the histograms
in Fig.2 and Fig.3 for the two simple but generic tests that we have 
explored here.  

Lessons learned in this study can be applied to
the monopole BAO analysis  (eg Eisenstein et al 2005, Sanchez et al 2009,  
Percival et al 2010, Kazin et al 2010a)  and the radial BAO in GCH 
(or the BAO in the 3-point function by Gaztanaga et al 2009).
Kazin et al (2010b) have argued that because $\omega$CDM  does not fit 
 the radial BAO data significantly better than a model with $\xi=0$ 
(null test),  the $H(z)$ measurements presented by CGH 
based on the location of the radial BAO peak can not be 
regarded as a detection. We have shown here that his argument
is not necessarily correct. %EG
If we apply such argument to
the monopole BAO measurements previously cited we would conclude
that we can not locate the peak position in current data because
according to Fig.2 there is no significant BAO detection.
But we have shown here that we can locate the BAO position with 
reasonable accuracy  even with data that fails the null BAO detection test. 
A  similar analysis was done with Monte Carlo mocks
in  GCH for the radial BAO position.
Even if the shape method gives low significance, 
the peak method can still be used to detect the position of the peak.

Tian et al 2010 reaches similar
conclusions for the radial BAO peak using different simulations.
They use a wavelet technique to detect the peak and asses the significance of the
detection, splitting SDSS into slices in various rotations. 
%A recent work by Tian et al 2010 reaches 
%equivalent conclusions applied also to the radial BAO peak using 
%simulations and a more elaborate peak finder.
%EG

Current BAO measurements can not yet be used
to select $\omega$CDM, but they can be used  to locate the BAO
position  (or other cosmological parameters)
if one assumes $\omega$CDM  or models which
produce similar clustering (and errors) to the ones in  $\omega$CDM.
\footnote{Data and mocks used in this paper, together with
covariance matrix and additional figures can be found in
http://www.ice.csic.es/mice/baodetection.}

\section*{Acknowledgements} 
 
We would like to thank Carlton Baugh, Pablo Fosalba, Ramon Miquel
and Ariel Sanchez for their
comments on earlier versions of this manuscript.
The MICE simulations have been developed at the MareNostrum
supercomputer (BSC-CNS) thanks to grants AECT-2006-2-0011 through
AECT-2010-1-0007. Data products have been stored at the Port
d'Informaci\'o Cient\'ifica (PIC). 
This work was partially supported by 
the Spanish Ministerio de Ciencia e Innovacion (MICINN), projects
AYA2009-13936, Consolider-Ingenio CSD2007- 00060 and
research project 2009-SGR-1398 from Generalitat de Catalunya.

\end{document}